%% file: eptcs-paper.tex
\documentclass[
  submission,
  copyright,
  creativecommons,
  sharealike,
  dvipsnames
]{eptcs}
\providecommand{\event}{TFPIE 2023} 

\usepackage{iftex}
\usepackage{acronym}
\usepackage{orcidlink}
\usepackage{tabularx}
\usepackage{subcaption}
\usepackage{algorithmicx}
\usepackage{algpseudocode}
\usepackage{algorithm}
\usepackage{amssymb}
\usepackage{amsmath}
\usepackage{listings}
\usepackage{lstautogobble}
\usepackage[capitalize]{cleveref}
\usepackage{pgf-pie}
\usepackage{xcolor}
\usepackage{pgfplots}
\usepackage{csquotes}
\usepackage[inline]{enumitem}
\usepackage{wrapfig}
\usepackage{colortbl}

\makeatletter
\@ifclasswith{eptcs}{adraft}{\usepackage{todonotes}}{\usepackage[disable]{todonotes}}
\makeatother

\ifpdf
  \usepackage{underscore}         
  \usepackage[T1]{fontenc}        
\else
  \usepackage{breakurl}           
\fi

\hypersetup{hidelinks}
\graphicspath{{img}}
\algrenewcommand\algorithmicrequire{\textbf{Input:}}
\algrenewcommand\algorithmicensure{\textbf{Output:}}
\definecolor{myOrange}{HTML}{E69F00}
\definecolor{myBlue}{HTML}{56B4E9}
\definecolor{myGreen}{HTML}{009E73}
\definecolor{myPink}{HTML}{CC79A7}

\author{
\orcidlink{0000-0001-5962-6583}
Ole Lübke\footnotemark
\quad
\orcidlink{0000-0002-8472-1483}
Konrad Fuger\footnotemark
\\
\orcidlink{0000-0002-5204-4713}
Fin Hendrik Bahnsen\footnotemark\textsuperscript{~~,}\footnote{F. H. Bahnsen was with the Institute of Embedded Systems, TUHH, when the presented work was created.}
\quad
Katrin Billerbeck\footnotemark
\quad
Sibylle Schupp\footnotemark[1]
\email{\{ole.luebke, k.fuger, fin.bahnsen, katrin.billerbeck, schupp\}@tuhh.de}
\institute{
\begin{minipage}{.49\textwidth}
  \centering
  \footnotemark[1]{}~~Institute for Software Systems\\
  \footnotemark[2]{}~~Institute of Communication Networks\\
  \footnotemark[5]{}~~Center for Teaching and Learning\\
  Hamburg University of Technology (TUHH)\\
  Hamburg, Germany
\end{minipage}
\begin{minipage}{.49\textwidth}
  \centering
  \footnotemark[3]{}~~Institute for Artificial Intelligence in Medicine\\
  University Medicine Essen\\
  Essen, Germany
\end{minipage}
}
}

\def\titlerunning{Computer Aided Design and Grading for an Electronic FP Exam}
\def\authorrunning{O. Lübke, K. Fuger, F. H. Bahnsen, K. Billerbeck, S. Schupp}

\newacro{FP}{functional programming}
\newacro{LO}{learning objective}
\newacro{CA}{constructive alignment}
\newacro{TUHH}{Hamburg University of Technology}
\newacro{YAPS}{\textit{Your Open Examination System for Activating and emPowering Students}}
\newacro{GHC}{Glasgow Haskell Compiler}
\newacro{RegEx}{regular expression}
\newacro{HTML}{Hypertext Markup Language}
\newacro{JSON}{JavaScript Object Notation}
\newacro{HTRSL}{Haskell Task \ac{RegEx} Specification Language}
\newacro{e-exam}{electronic exam}
\newacro{DAG}{directed acyclic graph}
\newacro{CYP}{\textit{Check Your Proof}}
\newacro{AST}{abstract syntax tree}
\title{Computer Aided Design and Grading for an Electronic Functional Programming Exam}

\begin{document}
\maketitle

\input{abstract}

\input{sections/intro}
\acresetall
\input{sections/relwork}

\acresetall
\input{sections/background}

\acresetall
\input{sections/analysis}

\acresetall
\input{sections/realization}

\acresetall
\input{sections/evaluation}

\acresetall
\input{sections/summary}

\bibliographystyle{eptcs}
\bibliography{tfpie23}

\input{sections/appendix}

\end{document}

%% file: abstract.tex
\begin{abstract}
	Electronic exams (e-exams) have the potential to substantially reduce the effort required for conducting an exam
	through automation.
	Yet, care must be taken to sacrifice neither task complexity nor constructive alignment nor grading fairness in favor of
	automation.
	To advance automation in the design and fair grading of (functional programming) e-exams, we introduce the following:
	A novel algorithm to check \textit{Proof Puzzles} based on finding correct sequences of proof lines
	that improves fairness compared to an existing, edit distance-based algorithm;
	an open-source static analysis tool to check source code for task relevant features by traversing the abstract syntax tree;
	a higher-level language and open-source tool to specify regular expressions that makes creating complex regular
	expressions less error-prone.
	Our findings are embedded in a complete experience report on transforming a paper exam to an e-exam.
	We evaluated the resulting e-exam by analyzing the degree of automation in the grading process, asking students for
	their opinion, and critically reviewing our own experiences.
	Almost all tasks can be graded automatically at least in part (correct solutions can almost always be detected as such),
	the students agree that an e-exam is a fitting examination format for the course but are split on how well they
	can express their thoughts compared to a paper exam,
	and examiners enjoy a more time-efficient grading process while the point distribution in the exam results was almost
  exactly the same compared to a paper exam.

	\vspace{0.5em}\noindent\textbf{Keywords:} Electronic Examination, Automated Grading, Constructive Alignment
\end{abstract}

%% file: sections/intro.tex
\section{Introduction}\label{sec:intro}

A strong argument for \acp{e-exam} is their high potential for automation,
which can decrease the effort of conducting an exam, especially during grading.
Yet, care must be taken to sacrifice neither task complexity \cite{krathwohl2002} nor \ac{CA} \cite{biggs1996} in favor
of automation.
For example, an exam consisting only of multiple choice tasks would be easy to grade automatically,
but it would also be impossible to assess the ability of students to \emph{create} a solution rather than
\emph{select} a given one.
On the other hand, it is possible to integrate arbitrary program logic into an \ac{e-exam},
which does not only help with automating grading
(checking the correct answer through means ranging from simple equality checks,
over specialized, self-written algorithms, to possibly even artificial intelligence \cite{schneider2022}),
but can also be used for task design
(e.g., randomization to impede cheating, sensibly showing or hiding input fields depending on previously-given answers,
execution of student code to allow students to test their answers in a controlled way).
Following the recent introduction of large scale electronic examinations \cite{sitzmann2022}
and the development of the extensible \ac{YAPS} \cite{bahnsen2021} at \ac{TUHH},
we investigate in this paper how to leverage that opportunity to improve the examination in our \ac{FP} course.
We provide a complete tour of how we successfully designed an \ac{e-exam} for the \ac{FP}
course at \ac{TUHH}, both from a technological and an educational perspective.

As a first step we analyzed our pre-\ac{e-exam} \ac{FP} courses and past exams to identify potentials for automation and
improvement.
The analysis was driven by the \acp{LO} of the course and how past exam tasks aligned with them.
We found that \acp{RegEx} are a good candidate for automatically checking answers that
consist of short source code snippets.
More elaborate programming tasks could be evaluated using software tests.
One particular \ac{LO} (Students are able to interpret compiler warnings and errors) had low coverage,
but in the \ac{e-exam} we could alleviate that by integrating live compiler feedback.
Overall, we found that many of our existing tasks could reasonably be transformed to an \ac{e-exam} version.

We implemented our ideas in YAPS, extending the system where necessary.
Because we teach \ac{FP} using the example of Haskell but \ac{YAPS} only offered a C/C++ compiler,
we extended it with a common template for Haskell programming tasks.
The template facilitates the quick creation of new tasks,
and features a tool to analyze student code for task-relevant features (e.g., usage of pattern matching)
that we developed specifically for the \ac{e-exam}.
Another extension is checking answers with \acp{RegEx}.
We found that crafting \acp{RegEx} that are flexible enough to accept all valid solutions,
but also strict enough to reject any wrong answers, is a challenging and time-consuming process.
Therefore, we devised a small, high-level specification language for \acp{RegEx} tailored to common patterns found in
Haskell code, and a tool that compiles these specifications to actual \acp{RegEx}.
Furthermore, \ac{YAPS} features a way for students to enter proofs (\textit{Proof Puzzles})
that are then graded by an algorithm based on edit distance between correct solution and given answer.
This algorithm produces results that differ from our own, manual evaluation.
Consequently, we developed a new algorithm based on the length of correct proof sequences that is in line with
our judgment.
Finally, while a paper exam is very flexible in terms of input (i.e., anything that can be written or drawn),
an \ac{e-exam} is not because it requires specific input in specific input fields
(e.g., a numeric input field only accepts numbers).
Because we did not want to take that flexibility away from students,
we extended YAPS with a comment field for each task that students can freely use to their liking,
e.g., for taking notes, or noting assumptions they made in case they deem the task ambiguous.

For evaluation, we manually analyzed the degree of automation of the \ac{e-exam} (focusing on automated grading),
and asked the students that participated in the exam for their opinion.
Additionally, we report our own experiences and reflect on sources of errors in automated grading.
Grading of almost all tasks features at least some automated elements; some tasks are graded fully automated.
For almost all tasks correct answers can be awarded the full amount of points automatically.
Yet, awarding a sensible amount of points for partially correct solutions remains a challenge in most cases.
For us, the most common mistakes in automated grading were related to tasks checked by \acp{RegEx} and programming
tasks.
Currently, experience and manual testing are the best defense against errors, but our \ac{RegEx} creation tool
is a first step towards more (but less error-prone) automation in that regard as well.
The results from the student poll are encouraging,
as most students state that an \ac{e-exam} is the right format for the \ac{FP} course.
Yet, they are split on how well they can express their thoughts in comparison to a paper exam.

Altogether, we contribute the following to advancing computer aided design and grading for \ac{FP} \acp{e-exam}:
\begin{itemize}
	\item An algorithm to check \textit{Proof Puzzles} (cf. \Cref{sec:proof-algo}).
	\item A small static analysis tool to check Haskell code for exam task relevant features (cf. \Cref{sec:analysis-tool}).
	\item A language to specify \acp{RegEx} on Haskell snippets (cf. \Cref{sec:regex-gen}).
\end{itemize}
In addition, this paper is a complete experience report which highlights challenges when moving from a pen-and-paper to
an \ac{e-exam}, and provides answers to the following issues:
\begin{itemize}
	\item\label{challenge:quality}\textit{How can exam quality be ensured?}
	By analysis of the \acp{LO} of each task.
	The \ac{e-exam} should assess the same \acp{LO} as the paper exam (cf. \Cref{sec:analysis}).
	\item\label{challenge:fairness}\textit{How can fair grading be ensured?}
	By critically examining results from automated grading (cf. \Cref{sec:proof-algo}),
	giving students room to express their ideas (cf. \Cref{sec:comment-field}),
	and (for now) resorting to manual assessment if necessary (cf. \Cref{sec:plan,sec:autograding}).
	\item\label{challenge:progtasks}\textit{How can programming tasks with compiler support be realized in a way that
		feels natural to students, does not leak secret information about the task, and facilitates automation?}
	By employing a reusable template where the \texttt{main} function cannot be edited by students
	(cf. \Cref{sec:compiler}).
	\item\label{challenge:restrictions}\textit{How can certain restrictions in programming tasks be checked?}
	By syntactic analysis of the submitted code (cf. \Cref{sec:analysis-tool}).
	\item\label{challenge:ambiguity}\textit{How can an empty answer be distinguished from the absence of an answer?}
	By combining ambiguous input fields with a suitable yes/no question (cf. \Cref{sec:regexs}).
	\item\label{challenge:regexs}\textit{How can the absence of errors be ensured in complex \acp{RegEx} that are used to
    check answers?}
	By employing a higher-level, domain-specific language that compiles to \acp{RegEx} (cf. \Cref{sec:regex-gen}).
\end{itemize}

The paper is organized as follows:
\Cref{sec:relwork} discusses related work,
and \Cref{sec:background} introduces \acf{CA} and the \ac{YAPS} examination software.
\Cref{sec:analysis,sec:realization,sec:evaluation}
contain the above-mentioned analysis, realization, and evaluation respectively.
\Cref{sec:summary} summarizes and provides an outlook on future work.

%% file: sections/relwork.tex
\section{Related Work}\label{sec:relwork}

While electronic assessment and automated grading have been an active research area for decades
\cite{hollingsworth1960,paiva2022},
the design of entire \acp{e-exam} seems to get less attention.
The report on introducing \acp{e-exam} in two (Java-based) programming courses by Rajala et al.\ \cite{rajala2016}
appears closest to our work.
They also establish a categorization of tasks, focus on complete exam designs,
and use questions similar to ours to collect student feedback.
The most notable difference is the method to ensure exam quality: the authors asked two (otherwise uninvolved) other
researchers for their opinion.
Instead, we suggest to make sure all \acp{LO} of the course are assessed in the exam through analysing each task.
For additional quality assurance, both methods could be combined.
Other differences include that Rajala et al.\
require source code submitted in the exam to compile (which we decided against),
check correctness by comparing the program output (instead, we use property tests), and they
neither restrict which code can be executed, nor the input language.
These differences can likely be attributed to different \acp{LO} in the respective courses and the different
programming languages and paradigms.

Bloom's Taxonomy \cite{krathwohl2002} is a tool widely used to assess task complexity,
and can therefore also be used to assess an entire exam.
Still, Sheard et al.\ developed a scheme for classifying exam tasks using seven different features,
targeted specifically at introductory computer science courses \cite{sheard2011}.
Our analysis is different in that its aim is not to characterize a single exam, but to make sure exam quality did not
degrade from one to the next exam.
It uses the \acp{LO} covered by each task as the only feature to make sure all tasks are still constructively aligned
\cite{biggs1996}.

An older study focussing on the transition from paper to \ac{e-exam} has been conducted by Stergiopoulos et al.\
for a course in \textit{Electronic Physics} \cite{stergiopoulos2006}.
However, the set of investigated task types is limited to yes/no questions,
multiple choice, and calculations (with numeric input that must lie in a specific range).

Regarding challenges when introducing \acp{e-exam}, Kuikka et al.\ surveyed educators at Turku University of Applied
Sciences for which requirements for introducing \acp{e-exam} they see \cite{kuikka2014}.
In contrast, we focus on challenges when those requirements are met and an actual \ac{e-exam} is created.

During the COVID-19 pandemic, educators worldwide had to move their classes online and shared their experiences.
Often, \acp{e-exam} are also part of these reports,
but with varying level of detail regarding task design and automated grading.
Loftsson and Matthíasdóttir report on how they combined different online teaching tools to transform a first semester
(Python-based) programming course \cite{loftsson2021}.
They evaluate the changes in depth based on student surveys and exam results, but, except for the employed tools,
no details on task design and automated grading were given.
Kappelmann et al.\ focus on how students can be engaged in online teaching in an introductory (Haskell-based) \ac{FP}
course, and also provide details on task designs and automated grading \cite{kappelmann2022}.
For checking programming tasks, they also use property testing,
and add unit tests as well as a novel \texttt{IO} mocking library.
The latter can be seen as another way of controlling the execution of student code (we prevent \texttt{main} from being
changed), yet there is no mention of restricting the input language for certain tasks.
For checking proof tasks, they include an actual proof checker instead of \textit{Proof Puzzles}.

The \textit{Proof Puzzles} in \ac{YAPS} are a similar to \textit{Proof Blocks} \cite{poulsen2022a,poulsen2022}:
students are presented with building blocks containing lines of the proof as well as distractors,
and are supposed to construct the proof via drag-and-drop.
In \textit{Proof Blocks}, the proof structure is encoded as a \ac{DAG},
and the points to award for a given solution are calculated using an edit distance measure.
\textit{Proof Puzzles} also rely on edit distance for awarding partial points.
We found that this algorithm may produce unfair results and consequently developed a new one.
One the one hand, our algorithm is based on pre-defined correct block sequences which can be seen as non-branching \acp{DAG}.
On the other hand, we use a different mechanism for awarding partial points.
Generally, points are awarded for each correct block in a sequence.
Apart from the sequences, entry points are specified to allow for reentering the correct proof,
which allows for a fair distribution of points even when some mistakes were made.
An alternative to such block-based automatic proof grading is using a theorem prover \cite{kappelmann2022,jacobsen2023}.
For instance, Kappelmann et al.\ employed \ac{CYP} \footnote{\url{https://github.com/noschinl/cyp}} in an \ac{FP} exam
\cite{kappelmann2022}.
The input language of \ac{CYP} is intentionally close to Haskell, and it supports checking equational reasoning,
proofs by structural induction, proofs by extensionality, case analysis, and computation induction.
For the exam, they did not require the students to strictly adhere to the syntax of \ac{CYP}.
Still, similar to many tasks in our \ac{e-exam}, correct solutions, if given in the correct syntax, can automatically
be verified as such.
Additionally, establishing a common syntax for proofs in the course also facilitated the manual grading of partially
correct submissions.
Another example of a theorem prover used in an exam is given by Jacobsen et al.,
who integrated Isabelle/HOL \cite{nipkow2002} in a course on automated reasoning and report similar experiences with
respect to grading.
In comparison to theorem provers, \textit{Proof Puzzles} or \textit{Blocks} allow for fully automated proof grading,
but they cannot reason about the proof itself.
Instead, they rely on pre-defined structural information, such as the \ac{DAG} representation.
Yet, the two options do not necessarily exclude each other, as shown by McCartin-Lim et al.\ \cite{mccartin-lim2018}.
They developed a graphical, graph-based user interface where proof assumptions and assertions need to be selected
by students and dragged to the graph pane, where they can be connected with directed edges.
In the background, a theorem prover checks for each each assertion whether its proof is complete, and the outcome
is indicated visually.

Automatically grading programming tasks is a vast field that has recently been reviewed by Paiva et al.\
\cite{paiva2022}.
While it is often easy to determine whether a given program is correct through traditional testing,
it is much harder to assign a reduced amount of points to partially correct solutions.
Our approach is to execute multiple tests and use a set of rules that map test results to points.
However, manual inspection is still required for incorrect programs for two reasons:
Programs that do not compile cannot be evaluated this way (but we also want to grade those),
and there are cases where the assigned amount of points is too low because the tests did not cover a detail that we do
want to award points for.
Static analysis is a more formal way to measure the difference between a given and correct solution based on a graph
representation derived from the \ac{AST} and its usefulness has been shown in numerous studies \cite{paiva2022}.
We also use static analysis in the automated grading process, but not for assigning reduced points.
Instead, it is employed to check whether a given solution can be a viable solution at all.
Some of our programming tasks restrict the input language by forbidding/requiring the use of certain language features
or functions.
Since, to our knowledge, no tool exists that reports on the functions and language features used in a Haskell program,
we devised our own.

For grading smaller tasks that do not require writing a full program or function, we use \acp{RegEx}.
\acp{RegEx} are a known tool for automated assessment, with applications such as parsing output of student programs
\cite{morris2003,pieterse2013}, extracting information from student code \cite{morris2003,ureelii2019},
or even grading free-form text answers \cite{kadupitiya2016}.
In this work, we are mostly concerned with generating correct \acp{RegEx} that represent valid solutions to a given task
because writing complex \acp{RegEx} that accept different answers is tedious and error-prone.
Existing work on ensuring the correctness of \acp{RegEx} has recently been reviewed by Zheng et al.\ \cite{zheng2021}.
While there are numerous approaches to testing or even verifying existing \acp{RegEx}, methods for creating \acp{RegEx}
are mostly concerned with learning them either from examples or from a natural language description.
Another line of research is concerned with making \acp{RegEx} easier to understand for humans,
e.g., by abstraction \cite{erwig2012}.
Our approach to generate \acp{RegEx} can be seen as a reverse abstraction process.
We use a more abstract language that is compiled to \acp{RegEx} instead of abstracting \acp{RegEx}.

%% file: sections/background.tex
\section{Constructive Alignment \& YAPS}\label{sec:background}


The concept of \emph{Constructive Alignment} by Biggs \cite{biggs1996} is related to constructivist learning theories,
i.e., knowledge is seen to be \emph{constructed} by each learner, as opposed to being \emph{transferred} from teacher
to learner.
Thus, in order to facilitate successful learning a learning environment must be designed that allows the desired
\acp{LO} to be achieved through appropriate learning activities.
The key to this is the definition of \acp{LO} at an appropriate level as well as the design of matching
examination tasks.
This is because, according to the principle of \emph{testing drives learning},
the cognitive level of the examination tasks also determines the learning activities of the students.
Thus, if what is to be learned is not reflected in the examination tasks,
well-intentioned motivational teaching often fails due to a lack of student engagement.

To achieve higher cognitive learning objectives in an e-exam in engineering science,
a technical environment is needed that offers much more than the construction of multiple choice tasks.
Students should be able to analyze presented issues, to develop and justify independent solutions,
and not just select predetermined answers.
For such demanding examination scenarios that assess deeper understanding \cite{krathwohl2002},
the examination system \ac{YAPS} was developed at \ac{TUHH},
which allows for a more creative and technically appropriate construction of examination tasks \cite{bahnsen2021}.

Several key design decisions were made in the development of YAPS:
It is licensed open source and is cost-efficient for the university with regard to the required hardware resources.
In Germany, there is no real alternative to self-hosting of the examination system used for reasons of data protection.
This criterion alone excludes many candidates for other examination software.
\ac{YAPS} is characterized by a contactless and state-based operating concept that maps the testing procedure of \ac{TUHH},
i.e., steps like having each student confirm they are fit to take the exam, checking student ID cards and
whether they are actually registered, are supported by \ac{YAPS} and can be performed while keeping a distance
between each other (which was especially important during the COVID-19 pandemic, when exams at \ac{TUHH}
were still conducted in person).
Finally, \ac{YAPS} is designed to be extensible and built with well-known technologies (e.g., Typescript, Docker).

Other, similar softwares such as Autolab\footnote{\url{https://github.com/autolab/Autolab}} \cite{milojicic2011},
INGInious\footnote{\url{https://github.com/UCL-INGI/INGInious}} \cite{derval2015} or
ArTEMiS\footnote{\url{https://github.com/ls1intum/Artemis}} \cite{krusche2018} can also be self-hosted,
are extensible, and the latter even supports Haskell out of the box.
However, \ac{YAPS} is already available at \ac{TUHH} and well-integrated with the processes attached to conducting
an exam.
Its lack of support for Haskell can easily be remedied through its extensibility.

%% file: sections/analysis.tex
\section{Analysis of our Pre-E-Exam FP Course}\label{sec:analysis}

Changes to the examination of a course should be reflected in the teaching activities, and vice versa.
This ensures that the teaching activities prepare the students to take the exam,
and that the exam assesses the knowledge and skills taught in the course.
To be able to review this mutual dependency,
in this section we analyze our \ac{FP} course and exams before the \ac{e-exam}.

First, through describing and listing the teaching activities of the course, we capture its current state to identify
any exam-related parts that may need to be updated.
Then, we examine previous exams with regard to \ac{CA} and task categories to identify potentials for automation and
improvement of task design.
Finally, having established candidates for change in both the teaching activities and in the exam,
we devise a plan to reach our goal: an updated \ac{FP} course with an \ac{e-exam} that reduces effort through automation
without reducing exam quality.

\subsection{Course Description}

\begin{table}
	\caption{\Aclp{LO} of our \ac{FP} course as specified in the module handbook \cite{hamburguniversityoftechnology2022}}\label{tab:lo}
	\centering
	\begin{tabularx}{\textwidth}{|l|X|}\hline
		\multicolumn{2}{|c|}{Knowledge-based}                                                                                                                        \\ \hline
		$K_1$ & Students apply principles, constructs, and simple design techniques of \ac{FP}.                                                                      \\
		$K_2$ & Students demonstrate ability to read Haskell programs, and explain Haskell syntax.                                                                   \\
		$K_3$ & Students interpret warnings and find errors in programs.                                                                                             \\
		$K_4$ & Students apply fundamental data structures, data types, and type constructors                                                                        \\
		$K_5$ & Students employ strategies for unit tests of functions and simple proof techniques for partial and total correctness.                                \\
		$K_6$ & Students distinguish laziness from other evaluation strategies.                                                                                      \\ \hline
		\multicolumn{2}{|c|}{Skill-based}                                                                                                                            \\ \hline
		$S_1$ & Students break a natural language description down in parts amenable to a formal specification and develop a functional program in a structured way. \\
		$S_2$ & Students assess different language constructs, make conscious selections both at specification and implementation level, and justify their choice.   \\
		$S_3$ & Students analyze given programs and rewrite them in a controlled way.                                                                                \\
		$S_4$ & Students design and implement unit tests and can assess the quality of their tests.                                                                  \\
		$S_5$ & Students argue for correctness of their program.                                                                                                     \\\hline
	\end{tabularx}
\end{table}

The course is based on the textbook \enquote{Programming in Haskell} by Graham Hutton \cite{hutton2016} and mainly
targeted at first semester computer science students.
The teaching activities of the course are aligned with the \acp{LO} summarized in \cref{tab:lo}.
Overall, there are three weekly activities:
First, the lecture lays the foundation for the other activities.
To a limited extent, it also features practical elements like executing code examples and short in-class exercises.
Second, so-called programming labs provide an opportunity to get hands-on experience with \ac{FP}.
During the labs, students solve small programming tasks with the support of student tutors, as it is common in other
courses as well \cite{bieniusa2008}.
They can also help out each other and exchange different approaches towards solving the tasks.
At the end of each lab session, the tutors discuss the solutions with each student, making sure that misunderstandings
are detected and cleared up as early as possible.
Third, there are homework exercise sheets that, in contrast to the labs, are supposed to be solved alone.
They feature more complicated tasks that require and foster a deeper understanding.
To check their solutions for correctness, students upload their code files to an autograding system,
so they can get feedback anytime.
Usually the students need to provide a number of examples and tests for each function,
and the given feedback reports on the correctness of functions, examples, and tests alike.
For failing tests also the test input, expected and actual output are given.
Finally, the solutions are discussed in a dedicated lecture hall exercise session.

During the last month of the lecture period, we introduce a few changes to what is described above,
to help the students prepare for the exam.
The last two tasks of each programming lab are designed to revisit topics covered earlier
in the lecture period.
During the last lab session, students are presented with the opportunity to solve an old exam,
so they get an idea of what to expect during the real one.

\subsection{Existing Exams}

The exam is usually divided into eight main tasks with subtasks,
where each of the main tasks is dedicated to a certain topic from the lecture:
\begin{enumerate*}
	\item Types and Type Classes
	\item List Comprehension
	\item Pattern Matching
	\item Recursion
	\item Higher-order Functions
	\item User-defined Types
	\item Evaluation
	\item Reasoning and Testing.
\end{enumerate*}
To ensure we preserve \ac{CA} while transforming the exam, and to get an overview of what types of task we have,
we analyzed the tasks of an exemplary old exam.
The results are summarized in \cref{tab:oldTasks}.
During the analysis we noticed that there are no warning messages in the exam, so the first part of $K_3$ is not
checked at all.

  \begin{wraptable}{R}{.55\textwidth}
	\caption{\Aclp{LO} and types of existing exam tasks}\label{tab:oldTasks}
	\centering
	\begin{tabular}{l|l|l}
		Task  & \Acp{LO}                                 & Type                   \\ \hline \hline
		1a    & $K_1$, $K_2$, $K_4$                      & snippet                \\
		1b    & $K_1$, $K_2$, $K_4$                      & multiple choice        \\ \hline
		2a    & $K_1$, $K_2$, $K_3$                      & single choice, snippet \\
		2b    & $K_1$, $K_2$, $S_2$, $S_3$               & code                   \\ \hline
		3a    & $K_1$, $K_2$, $K_3$, $K_4$               & single choice, snippet \\
		3b    & $K_1$, $K_2$, $K_4$, $S_3$               & code                   \\ \hline
		4a    & $K_1$, $K_2$                             & text                   \\
		4b    & $K_1$, $K_2$, $S_1$, $S_2$, $S_3$        & code                   \\ \hline
		5a    & $K_1$, $K_2$                             & snippet                \\
		5b    & $K_1$, $K_2$                             & text                   \\
		5c    & $K_1$, $K_4$, $S_1$, $S_2$               & code                   \\ \hline
		6a i  & $K_1$, $K_2$, $K_4$, $S_1$               & snippet                \\
		6a ii & $K_1$, $K_2$, $K_3$, $K_4$               & single choice          \\
		6b    & $K_1$, $K_4$, $S_1$                      & code                   \\ \hline
		7a    & $K_1$, $K_2$, $K_6$                      & text                   \\
		7b    & $K_1$, $K_2$, $K_6$                      & snippet                \\
		7c    & $K_1$, $K_2$, $K_6$                      & text                   \\ \hline
		8a    & $K_1$, $K_2$, $K_5$, $S_1$, $S_2$, $S_4$ & code                   \\
		8b    & $K_1$, $K_2$, $K_5$, $S_5$               & text
	\end{tabular}
\end{wraptable}

Regarding the types of tasks, we found that all tasks can be presented in one of the following five categories:
\begin{enumerate*}
	\item Single choice: Select one option from two or more.
	\item Multiple choice: Select zero or more options from two or more.
	\item Snippet: Write a short piece of source code, e.g., a type, an expression, or even just a number.
	\item Code: Write a larger piece of source code, e.g., a complete function or user-defined type.
	\item Text: Write a text, e.g., some explanation or justification.
\end{enumerate*}

To make sure the \ac{e-exam} assesses the same \acp{LO} as we did previously, we took the arguably most simple approach
to design it: \enquote{translating} each task directly.
With the first two categories we already use task types that can directly be integrated into an \ac{e-exam}.
Integrating snippet and code tasks is more difficult, but the answers must follow the rules of Haskell by design,
so at least partially automated assessment is possible for them.
Text answers are comparatively unstructured, and students may answer in either English or German,
which makes those tasks ineligible for automated assessment.

\subsection{Consequences}\label{sec:plan}

The analysis of the teaching activities showed that the later lab sessions with recapitulation tasks are suited best
for familiarizing the students with the \ac{e-exam} format.
Therefore, we decided to move those tasks to \ac{YAPS} instead of using text editor and terminal as usual.
In addition, the exemplary exam that students can solve during the last lab appointment is also in electronic form.
This way, all students have the opportunity to learn and try out how to use \ac{YAPS},
and we can also test new ideas in a comparatively risk-free environment.
Furthermore, our tutors are not only there to help out with any issues that may arise,
but also to collect valuable feedback directly from the students, which we could not get any other way
(especially observations on how the system is used and insights from student-to-student/tutor conversations).
Through this feedback we can detect problems (e.g., task designs that are difficult to navigate)
early and prevent them from occurring in the real exam.

Regarding the exam itself, we found ways to transform each of the task categories we identified to an
\ac{e-exam}.
Single and multiple choice tasks can naturally remain.
For snippet tasks we found that \acp{RegEx} can in all cases be used to detect correct solutions,
yet a \ac{RegEx} match is a Boolean decision (i.e., either the given answer matches the \ac{RegEx} or not),
so awarding a reduced amount of points for partially correct solutions is not possible.
This is why we also allow for a manual assignment of points in case there was no match.
Alternatively, snippet tasks could be checked using a compiler.
While this does not solve the problem of grading partially correct solutions, it would reduce the workload because
no \acp{RegEx} need to be created.
On the other hand, \acp{RegEx} are arguably more flexible because they could also be used to detect common errors
(by additionally specifying multiple erroneous \acp{RegEx}), hence enabling awarding reduced amounts of points.
Tasks that require writing a larger amount of source code can also remain,
but additional work is required for the automated assessment here as well.
In \ac{YAPS}, programming tasks are worked on in an integrated code editor with syntax highlighting and the possibility
to compile and execute the code, where the output of both compilation and execution is shown to the students.
For automated assessment we mainly employ randomized property tests via QuickCheck \cite{claessen2000} and
a new static analysis tool that provides information on used language features and functions.
Welcome side effects of integrating a code editor are that programming tasks can now be executed in a more familiar way,
and through error and warning messages from the compiler, \ac{LO} $K_3$ is now fully included,
while in the paper exam it was not.
One important decision to make with automated programming tasks is whether only compiling code is accepted as an answer,
or whether (partial) points are awarded also for non-compiling programs.
We decided for the latter because we believe that a solution that is correct except for, e.g., a single
wrongly-indented line, should still be worth points.
Still, non-compiling code cannot be tested, so this reduces the degree of automation.
Plain text tasks are the hardest to automate and there is no general way to do this.
Therefore, we examined each task individually and found that there actually is a way to transform many of the text tasks
in our exam in a similar way.
We often ask students to make a decision and then justify it,
so at least the decision part can be modeled as a single or multiple choice task,
and only the reasoning is left for manual evaluation.
Tasks in which we ask students to prove, e.g., a certain property of a function are a special case though.
Here we make use of the proof puzzle task type available in \ac{YAPS}, with a new evaluation algorithm tailored to our
needs.

%% file: sections/realization.tex
\section{Realization}\label{sec:realization}

In the following we explain in detail how we realized the plan derived from
the course analysis.
First, we focus on the extensions to \ac{YAPS}: Haskell compiler integration, \ac{RegEx} tasks, proof puzzle evaluation,
and the addition of a comment field for all tasks.
Then, we introduce the code analysis tool for Haskell programming tasks and the \ac{RegEx} generator program,
together with its higher-level \ac{RegEx} specification language.

\subsection{Haskell Compiler Integration}\label{sec:compiler}

To facilitate creating new programming tasks, we devised a template that provides a common layout and structure,
so only task-specific text and code needs to be added using the following workflow:
\begin{enumerate*}
	\item Write the task description into \textit{exercise.html}.
	      This is the page that students see when selecting the task.
	\item Write a short \texttt{main} function that executes the student code into \textit{main.hs}.
	      Students cannot modify this file, hence students cannot execute arbitrary code.
        They are limited to the functions evaluated in \texttt{main}, but we also control the types of these functions.
        Most importantly, these functions are never \texttt{IO} actions in our tasks,
        so we can leverage the type system of Haskell to prevent students from doing anything dangerous
        (the code is additionally isolated using containerization).
	\item Write any task-related code or comments into \textit{functions.hs}.
	      This is the file in which students implement their solutions.
	      It is a good idea to summarize the task in a comment here, so students do not have to switch back and forth
	      between the task description and their solution.
	\item Write tests for the student code into \textit{main\_test.hs}.
	      Students cannot see this file.
	\item Write the code to execute the tests, parse and interpret their output, and report the results back to \ac{YAPS}
	      into \textit{evaluate.py}.
	      Students cannot see this file.
\end{enumerate*}

This structure works for almost all tasks without modification, but is still flexible enough to allow for deviations.
One example of that are tasks where students need to write tests themselves.
Testing tests is more difficult than testing \enquote{normal} functions because we cannot write randomized property
tests for them.
Instead, we omit \textit{main\_test.hs} and provide two (or more) implementations of the function under test:
one that is correct, and one (or more) that is faulty.
The code from \textit{main.hs} is executed with the correct version during the exam.
During evaluation, we link the student test with each of the versions of the function under test.
Then we check whether the test passes for the correct version and fails for the faulty ones.

\subsection{Checking Answers with Regular Expressions}\label{sec:regexs}

\begin{figure}
	\begin{subfigure}{.49\textwidth}
		\includegraphics[keepaspectratio, width=\textwidth]{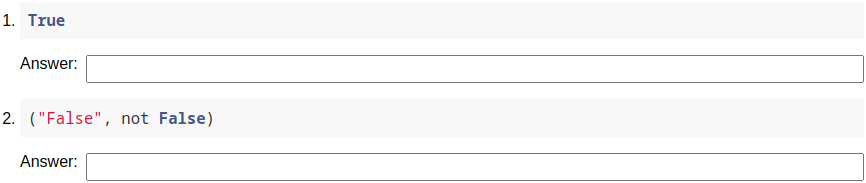}
		\caption{Exam view}\label{fig:regex-exam}
	\end{subfigure}
	\begin{subfigure}{.49\textwidth}
		\includegraphics[keepaspectratio, width=\textwidth]{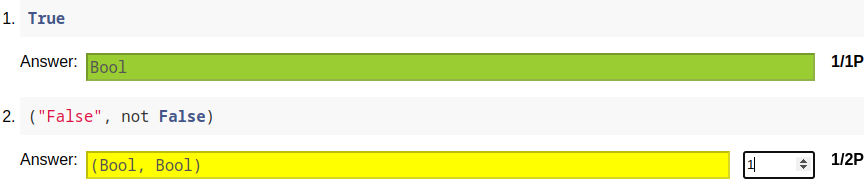}
		\caption{Evaluation view}\label{fig:regex-eval}
	\end{subfigure}
	\caption{\Acl{RegEx} task}
\end{figure}

\Cref{fig:regex-exam} shows for two instances of the \ac{RegEx} task type how they appear in the exam.
During the evaluation, the given text input is matched against the specified \ac{RegEx} and the results are displayed
to the examiner as in \cref{fig:regex-eval}.
The first answer is found to be correct.
The second answer is only partially correct,
so partial points are awarded through the numeric input field to the right.
This input field is hidden when the \ac{RegEx} matched to prevent any accidental modifications of the awarded points.

Still, there is one problem with that implementation: what if the correct answer is no answer?
Envision a task \enquote{Given a certain list comprehension, does it compile, and if so, what is the resulting list?}.
We usually design such tasks as a combination of a single choice (for the decision) and a \ac{RegEx} task
(for the list).
If the list comprehension is not valid, the correct answer would be to leave the list input field empty,
so we cannot distinguish between \enquote{no list} and \enquote{no answer.}
As a remedy, we allow \ac{RegEx} tasks to optionally depend on a single choice task.
The answer to the \ac{RegEx} part is only evaluated if an answer to the single choice task was given.
Additionally, the input field is hidden as soon as the student selects the negative option of the single choice part.
This resolves the ambiguity, because now we clearly know whether a task is answered.

\subsection{New Algorithm to Evaluate Proof Puzzles}\label{sec:proof-algo}

\begin{figure}
	\includegraphics[keepaspectratio, width=\textwidth]{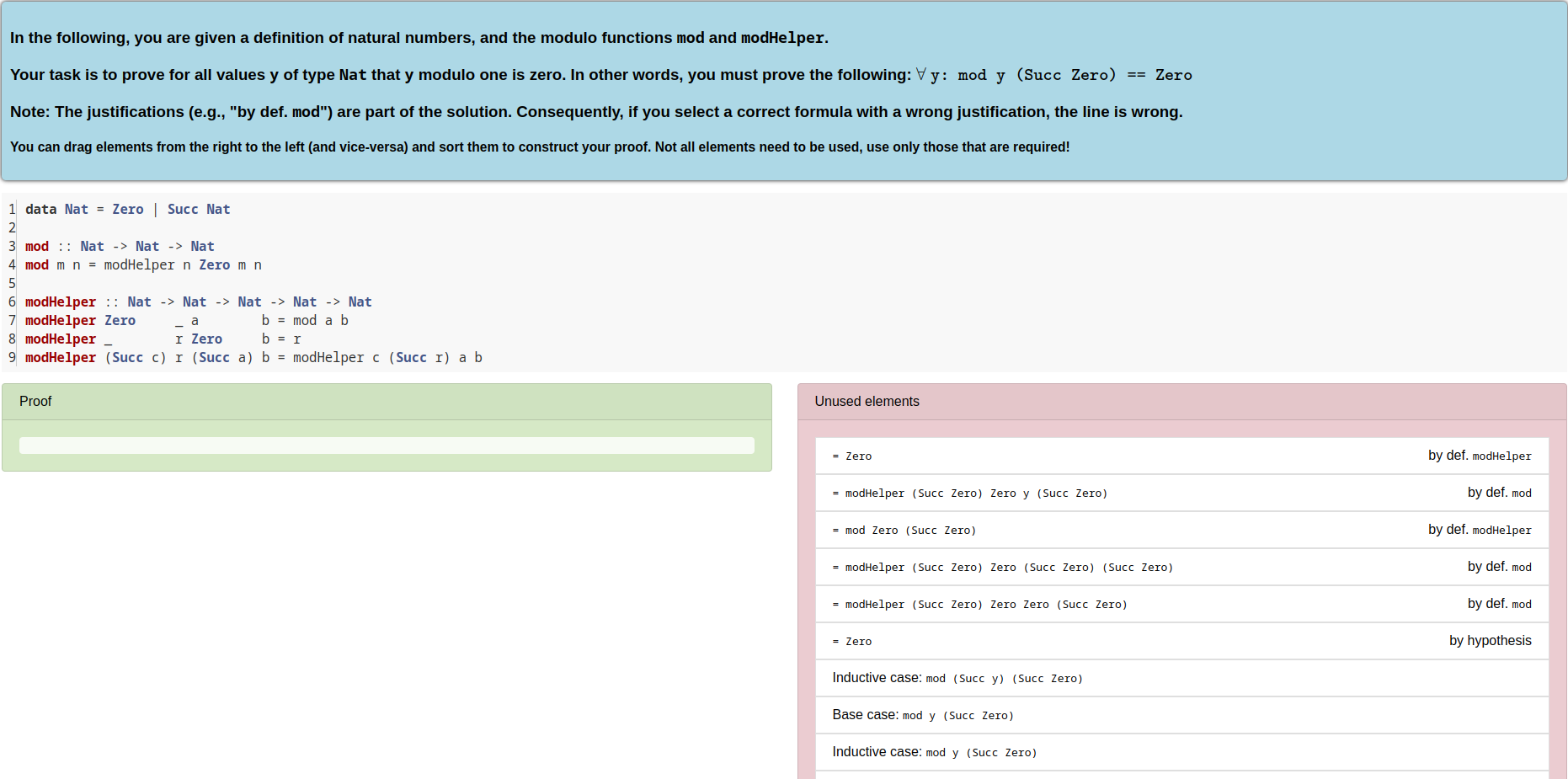}
	\caption{Proof puzzle task}\label{fig:proof}
\end{figure}

An example of the proof puzzle task type is given in \cref{fig:proof}.
Students can drag the elements from right to left (and vice versa) to construct their proof.
Such a task is defined as follows:
First, we list all the available items, and optionally assign a weight to them.
Then, we specify possible solutions as sequences of selected items, and assign each solution a number of points.
If a student produces one of these solutions exactly, they get the full amount of points.

For clarity, we introduce the following notation:
Let $\Sigma^*$ be the set of all (Unicode character) strings.
An \emph{item} $(t, w) \in \Sigma^* \times \mathbb{Q}$ consists of the displayed text $t$ and the associated weight $w$.
Let $\mathbb{S}_I$ be the set of all sequences of items in $I \subseteq \Sigma^* \times \mathbb{Q}$.
A \emph{solution} $(s, p) \in \mathbb{S}_I \times \mathbb{Q}$ of length $n$ consists of a sequence of items
$s = \iota_1, \dots, \iota_n$, and the associated number of points $p$.
We denote the $i^{th}$ item of $s$ as $\iota_i = (t_i, w_i)$.

In the original version of this task type, the automatic evaluation is based on the edit distance between the given
solution attempt and the specified solution.
The edit distance is defined as the weighted number of \emph{insert} and \emph{remove} operations required to transform
one sequence of items into another.
For an item $(\_, w)$, the cost of inserting/removing it is equal to $w$.
\Cref{alg:oldproof} (\nameref{sec:appendix}) shows the complete procedure for calculating the number of points to award for a solution attempt.
It can be summarized as follows:
For each possible solution, calculate the edit distance to the given solution attempt.
Then, subtract the edit distance from the maximum amount of points that can be awarded for that solution.
The resulting number of points is the maximum of these differences.

During testing, this algorithm produced reasonable results.
Yet, after conducting the exam,
we found that sometimes the resulting points were not in line with how we manually evaluated these tasks in the past,
and that sometimes the results were even unfair, e.g., in the following case
(cf. \Cref{fig:proof-old}, \nameref{sec:appendix}):
Student A has correctly identified the inductive step of the proof and the next step, and gets 0.5 points.
Student B has only correctly identified the base case and gets 2.0 points.
We could not find a different assignment of weights to the items that produces better results,
so we decided to devise a new algorithm tailored to our needs.

The new algorithm is based on finding correct sequences of items and awards points according to the item weights.
There is no subtraction of points, but sequences have clearly-defined entry points
(otherwise just using all items in a random order could yield the full amount of points).
As the entry points, we use the first items of the predefined solutions.
The new algorithm is given in \cref{alg:newproof} 
and replaces lines 3 and 4 in \cref{alg:oldproof} (\nameref{sec:appendix}).
It proceeds as follows:
First, initialize the result with 0 (line 11),
and look for the first entry point of a sequence in the given solution.
The subroutine to search for the next sequence start (line 1) takes the current indices in the lists of solution and
given answer items as input (both initially 0).
It iterates over the solution sequence, and checks whether the current item is an entry point,
and if it is also part of the given answer, a new synchronization point between solution and given answer is found.
The subroutine returns the indices of the common item in the solution and in the given answer.
If such a synchronization point cannot be found, the subroutine returns infinity for both indices,
which terminates the main loop (line 13).
The main loop iterates over the solution items as well,
checking for each item whether it is still in the sequence or breaks it.
If it is part of the sequence, its weight is added to the result,
and the indices in the lists of solution and answer items are incremented.
Otherwise, search for the next entry point and skip all items in between.

\Cref{alg:newproof} resolves the unfair grading explained above.
Student A now gets 1.5 points, while student B gets 1.0 (cf. \Cref{fig:proof-new}, \nameref{sec:appendix}).
In case of proof by induction tasks, we usually specify the first line of the base case and the first line of the
inductive step as entry points for sequences.

\begin{algorithm}
	\caption{New proof puzzle evaluation}\label{alg:newproof}
	\begin{algorithmic}[1]
		\Require $I \subset \Sigma^* \times \mathbb{Q}$ (items),
		$s \in \mathbb{S}_I \times \mathbb{Q}$ (solution),
		$a \in \mathbb{S}_I$ (solution attempt)
		\Ensure Amount of points to award for $a$
		\Function{FindNextSequenceStart}{$solIdx, ansIdx$}
		\For{$solIdx < |s|$}
		\If{$\Call{IsSequenceStart}{s_{solIdx}}\ \text{\textbf{and}}\ s_{solIdx} \in a$}
		\State \Return $(solIdx,\ \text{index of}\ s_{solIdx}\ \text{in}\ a)$
		\EndIf
		\State $solIdx \gets solIdx + 1$
		\EndFor
		\State \Return $(\infty, \infty)$
		\EndFunction
		\\
		\State $r \gets 0$
		\State $(solIdx, ansIdx) \gets \Call{FindNextSequenceStart}{0, 0}$
		\For{$solIdx < |s|\ \text{\textbf{and}}\ ansIdx < |s_a|$}
		\If{$s_{solIdx} = a_{ansIdx}$}
		\State $r \gets r + w_{solIdx}$
		\State $(solIdx, ansIdx) \gets (solIdx + 1, ansIdx + 1)$
		\Else
		\State $(solIdx, ansIdx) \gets \Call{FindNextSequenceStart}{solIdx, ansIdx}$
		\EndIf
		\EndFor
		\State \Return $r$
	\end{algorithmic}
\end{algorithm}

\subsection{Comment Field for Students}\label{sec:comment-field}

\input{sections/comment_field}


\subsection{Analyzing Student Code for Task-Relevant Features}\label{sec:analysis-tool}

Many of our programming tasks have certain restrictions, e.g., students must (not) use a certain language feature,
or are only allowed to use certain functions.
To be able to check these constraints automatically we devised a new
tool\footnote{\url{https://collaborating.tuhh.de/cda7728/check-hs-task-restrictions}} that itself is written in Haskell.
It takes a source code file as input, and outputs certain information on each function,
that is defined in the input file, in \ac{JSON}.
For example, with the well-known implementation of quicksort, given in \cref{lst:quicksort},
the program outputs the \ac{JSON} shown in \cref{lst:quicksort-out}.
For each function, the program provides the following information:
the name of the function, its arguments, called and locally declared functions,
and whether it uses pattern matching/guarded equations/list comprehensions/\texttt{case} expressions.
To extract that information,
we use the \texttt{haskell-src}\footnote{\url{https://hackage.haskell.org/package/haskell-src}} package to build and
traverse the \ac{AST} of the given source file.
The program itself is split into a library that provides the described functionality,
and an application that uses this library.
Altogether, the entire program consists of less than 500 lines of code of which most are used for pattern matching
on the constructors of the sum types that represent the \ac{AST} and advancing the search in depth-first manner.

\begin{lstlisting}[
      language=haskell,
      float,
      caption={Example of quicksort in Haskell\protect\footnotemark},
      label={lst:quicksort}
    ]
  quicksort :: Ord a => [a] -> [a]
  quicksort []     = []
  quicksort (p:xs) = (quicksort lesser) ++ [p] ++ (quicksort greater)
    where lesser  = filter (< p) xs
          greater = filter (>= p) xs
\end{lstlisting}
\footnotetext{\url{https://wiki.haskell.org/Introduction\#Quicksort_in_Haskell}}

\begin{lstlisting}[float,caption={Code analyzer output for \cref{lst:quicksort}},label={lst:quicksort-out}]
  { "functions": [{
      "name": "quicksort",
      "patMatch": true,
      "guards": false,
      "listComprehension": false,
      "hasIf": false,
      "hasCase": false,
      "args": [ "p", "xs" ],
      "calledFns": [ "quicksort", "++", "filter", "<", ">=" ],
      "declaredFns": [ ]
  }]}
\end{lstlisting}

The \texttt{haskell-src} library parses the given file to the \texttt{HsModule} type which features a list of declarations.
We filter those declarations for functions (\texttt{HsFunBind}) and constants  (\texttt{HsPatBind}),
and then check for the features and functions used.
The \texttt{HsFunBind} constructor holds the clauses of the function (\texttt{[HsMatch]}),
and the clauses provide access to guards (\texttt{HsGuardedRhss}),
as well as access to the patterns used in the left hand side.
Any pattern that is not a name (\texttt{HsPVar}) is considered a use of pattern matching.
Names, however, are added to the list of arguments (subpatterns are also considered, e.g., \texttt{(x:xs)} is
considered a use of pattern matching and adds the arguments \texttt{x} and \texttt{xs}).
To collect the remaining information, the right hand sides (\texttt{HsRhs}) and local declarations (\texttt{HsDecl})
are traversed similarly.

\subsection{Generating Flexible Regular Expressions}\label{sec:regex-gen}

To increase automation during exam creation we developed a second
tool\footnote{\url{https://collaborating.tuhh.de/cda7728/gen-hs-task-regexs}} in Haskell that facilitates writing the
\acp{RegEx} we need to automatically check many of the tasks.
It reads a custom, specialized specification language that we call \ac{HTRSL}, and outputs JavaScript compatible
\acp{RegEx}.
For an example, consider the input
\texttt{("Num" \textbackslash\textbackslash{} a) "=>" "[" a "]" "->" ["String" | "[" "Char" "]" ]}.
It represents the function type \texttt{(Num a) => [a] -> String} that can be written in different ways, e.g.,
omitting the parentheses, writing \texttt{[Char]} instead of \texttt{String},
or using a different name for the type variable \texttt{a}.
All of these variations need to be accepted by the generated \ac{RegEx}.
The result is shown in \cref{lst:regex-out} and illustrates how large these expressions can become.
\Acp{RegEx} of that size are difficult to understand for humans, so writing them by hand is rather error-prone.

\begin{lstlisting}[
      breakatwhitespace=false,
      float,
      caption={Example \ac{RegEx} generator output},
      label=lst:regex-out
    ]
  ^\s*(?:\((?=(?:[^\(\)]*)\)))?\s*Num\s*\s(?<a>[_a-z][_a-zA-Z0-9']*)\s*(?:(?<=\((?:[^\(\)]*))\))?\s*=>\s*\[\s*\k<a>\s*\]\s*->\s*(?:String\s*|\[\s*Char\s*\]\s*)$
\end{lstlisting}

Therefore, the \ac{HTRSL} provides an additional layer of abstraction tailored to common patterns in Haskell
expressions.
Its grammar is given in \cref{lst:htrsl-grammar} (\nameref{sec:appendix}) in labelled Backus-Naur form:
A \ac{HTRSL} file is a semicolon-separated list of specifications.
Each specification is a description.
A description is a list of description items.
Allowed items are:
\begin{itemize}
	\item Literals: Match the string given in double-quotes literally.
	\item Identifiers: Match any identifier (\texttt{[\_a-z][\_a-zA-Z0-9']*}).
	      If reused in the same description, only matches the same identifier that was found at the first occurrence.
	\item Mandatory whitespace: Match any whitespace (\texttt{\textbackslash s+}),
	      represented as \texttt{\textbackslash\textbackslash} in the description.
	\item Alternatives: Match any of the alternatives that are given in square brackets and separated by \texttt{|}.
	\item Optional parentheses: Match what is specified inside the parentheses, either surrounded by parentheses or not.
	      Cannot be nested.
\end{itemize}
We use the BNF Converter\footnote{\url{https://hackage.haskell.org/package/BNFC}} to generate a parser for the language.
This way, we obtain an abstract syntax tree that we can traverse to generate the desired \acp{RegEx}.

In the future, we plan to add support for automatically testing the generated \acp{RegEx}.
Currently, this is done manually by testing each \ac{RegEx} for some positive and negative examples.
This can be automated as well by adding the test strings to the specification,
so the tool can directly check whether the generated \ac{RegEx} matches all positive examples,
and rejects all negative ones.

%% file: sections/comment_field.tex
Another extension we added to the YAPS framework is a comment field for all exercises.
We frequently found in paper exams that students add additional thoughts about exercises beyond their answers.
Often these are explanations of how the exercise was understood or assumptions on which the answer of the student was
based.
While we aim to phrase all exercises in a way that neither an interpretation nor additional assumptions are necessary,
we still wanted to give students the option to express these so that their answers can be evaluated in the right
context.
Therefore, we added an all-purpose text field at the bottom of each exercise with an explanatory text describing its
purpose.
Whatever is written into this field persists in addition to all actual answers so that we can read it during the
correction of the exam.
Another use case we found for this text field is that students wrote notes for themselves during the exam that they
would ordinarily scribble in the margins of a paper exam.

%% file: sections/evaluation.tex
\section{Evaluation}\label{sec:evaluation}

For evaluating the \ac{e-exam} we aim to answer the following three questions:
What is the degree of automation?
Are students satisfied?
Are examiners (we) satisfied?
The first question is targeted at evaluating how far we have come with respect to our initial motivation:
reducing the effort of conducting an exam through automation.
We believe that, by design, we did not sacrifice \ac{CA}, yet \ac{CA} compliance comes at the cost of less automation,
and we wanted to know how high that cost actually is.
While reducing the effort of conducting the exam mostly benefits us, the examiners,
we also wanted to provide the students with a more comfortable and familiar way of taking an exam,
especially when it comes to programming tasks.
Finally, we reflect on our own experiences, from creating the exam to grading it, to answer the last question.

\subsection{Automated Grading}\label{sec:autograding}

\begin{wraptable}{R}{.57\textwidth}
	\centering
	\caption{Categorization of tasks by degree of automation}\label{tab:automation}
	\begin{tabular}{l|l|l}
		Task & Task Type              & Category             \\\hline\hline
		1a   & RegEx                  & Automated if correct \\
		1b   & Multiple Choice        & Fully automated      \\\hline
		2a   & Single choice + RegEx  & Automated if correct \\
		2b   & Programming            & Automated if correct \\\hline
		3a   & Single Choice + RegEx  & Automated if correct \\
		3b   & Programming            & Automated if correct \\\hline
		4a   & Single Choice + Text   & Partially automated  \\
		4b   & Programming            & Automated if correct \\\hline
		5a   & Programming            & Automated if correct \\
		5b   & Text                   & Not automated        \\
		5c   & Programming            & Automated if correct \\\hline
		6a   & Programming            & Automated if correct \\
		6b   & Multiple Choice        & Automated            \\
		6c   & Programming            & Automated if correct \\\hline
		7a   & Single Choice + Text   & Partially automated  \\
		7b   & RegEx                  & Automated if correct \\
		7c   & Multiple Choice + Text & Partially automated  \\\hline
		8a   & Programming            & Automated if correct \\
		8b   & Proof Puzzle           & Fully automated
	\end{tabular}
\end{wraptable}

To evaluate the degree of automation of the grading process,
we divide all tasks in the \ac{e-exam} into the following categories:
\begin{enumerate*}
	\item Fully automated: human intervention is not required in any case.
	\item Automated if correct: human intervention is only required if the given answer was incorrect
	      (to potentially award partial points)
	\item Partially automated: human intervention may even be necessary for correct answers.
	\item Not automated: human intervention is required in any case.
\end{enumerate*}
The results are given in \cref{tab:automation} for an example exam
(we do not rule out that other exams have, e.g., a task 1c).
They show a direct mapping between task type and category:
multiple/single choice or proof puzzle \textrightarrow{} fully automated,
\ac{RegEx} or Programming \textrightarrow{} automated if correct,
text \textrightarrow{} not automated.
When two task types are combined, the resulting degree of automation is the one that is worse.
Combinations of text tasks with one that is at least \enquote{automated if correct} result in a
\enquote{partially automated task}.
We see that most tasks are \enquote{automated if correct}, which is our second-best category.
Yet, just from the categorization we cannot determine how much this reduces the effort required for grading the exam
in practice.
Still, all tasks except for one feature at least some degree of automation, and overall that makes it likely to
reduce the required efforts.
In other words, the results can be summarized as follows:
Correct solutions can in most cases be awarded the full amount of points automatically,
but awarding partial points is difficult.

\subsection{Student View}\label{sec:eval-students}

To capture the view of the students, we conducted a short poll immediately after the exam.
When time for the exam was up, the exam browser automatically redirected the students to the poll website.
On the one hand this allowed us to capture the opinion of the students without delay and reduced external influence.
On the other hand, we thought that students may be unwilling to fill in an elaborate questionnaire just after finishing
an exam.
Therefore, we decided to only ask three questions and provide a text input field for additional feedback.

\begin{figure}
	\centering
	\begin{subfigure}{.3\textwidth}
		\begin{tikzpicture}
			\begin{axis}[ybar, width=\textwidth]
				\addplot coordinates {
						(1, 10)
						(2, 9)
						(3, 8)
						(4, 10)
						(5, 36)
					};
			\end{axis}
		\end{tikzpicture}
		\caption{Statement 1}\label{fig:poll-1}
	\end{subfigure}
	\begin{subfigure}{.3\textwidth}
		\begin{tikzpicture}
			\begin{axis}[ybar, width=\textwidth]
				\addplot coordinates {
						(1, 8)
						(2, 12)
						(3, 15)
						(4, 14)
						(5, 24)
					};
			\end{axis}
		\end{tikzpicture}
		\caption{Statement 2}\label{fig:poll-2}
	\end{subfigure}
	\begin{subfigure}{.3\textwidth}
		\begin{tikzpicture}
			\begin{axis}[ybar, width=\textwidth]
				\addplot coordinates {
						(1, 8)
						(2, 15)
						(3, 16)
						(4, 13)
						(5, 21)
					};
			\end{axis}
		\end{tikzpicture}
		\caption{Statement 3}\label{fig:poll-3}
	\end{subfigure}
	\caption{Poll results}\label{fig:poll}
	\textit{Rate the following statements from 1 to 5 (1 = fully disagree, 5 = fully agree)
		\begin{enumerate*}
			\item An \ac{e-exam} is a right format for the lecture \enquote{Functional Programming}.
			\item The programming tasks with compiler support feel like a natural way to answer programming tasks.
			\item In an \ac{e-exam}, I can express my thoughts as good as I could in a paper exam.
		\end{enumerate*}
	}
\end{figure}
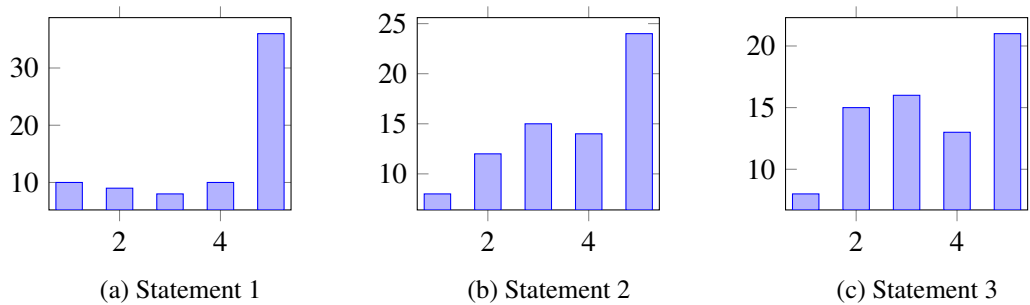

Of the 77 exam participants, 73 answered the poll.
The results are shown in \cref{fig:poll}.
\Cref{fig:poll-1} shows that the majority of the participants ($36 + 10 = 46 \equiv 63\%$) generally agrees that an
\ac{e-exam} is the right examination format for the course.
Most ($24 + 14 = 38 \equiv 52\%$) also found working with the integrated compiler natural,
yet comparatively many students are neutral on this question ($15 \equiv 21\%$), as shown in \cref{fig:poll-2}.
For the third question, \cref{fig:poll-3} shows that less than half of the participants ($21 + 13 = 34 \equiv 45\%$)
agreed that they can express their thoughts as they could in a paper exam.
Almost a third of the students ($8 + 15 = 23 \equiv 32\%$) disagrees.
29 students even provided additional feedback that ranges from constructive criticism to outright ecstasy
(\enquote{It was SIIIICKKK!!! Also perfect amount of time for the tasks. I finished in the last 5 seconds}).
Other critical remarks were:
\begin{itemize}
	\item \enquote{It's sometimes a little bit complicated for color-blind people to see whether I marked a task
		      already as green or it's still orange.}
	      In \ac{YAPS}, students can mark tasks green when they finished them, unseen tasks are gray, and tasks where some
	      input was given appear orange.
	      This choice of colors apparently is problematic for some and should be changed.
	\item \enquote{Maybe it would be better to make more friendly function names}
	      Sometimes the function names are a very short abbreviation that at first sight may appear as a random sequence
	      of letters.
	      Naming functions this way is a habit that stems from creating paper exams where horizontal space for code
	      is much more limited than in the \ac{YAPS} code editor, so we should consider using the additional space
	      and use more descriptive function names.
	\item \enquote{In my opinion it's nicer to write the proof (last task) as text instead of drag and drop, as the drag
		      and drop fields all look very similar, and it's more like a task to find the correct field, than to actually
		      think about the proof}
	      This is a hint that the proof puzzle may not be the optimal way to assess the ability of the students to
	      construct a proof, and we should investigate other ways.
	\item \enquote{Please get new computer mice. That is really necessary} (translated from German).
	      When designing the \ac{e-exam} we did not think about hardware,
	      apart from how the screen size impacts the visual presentation of the exam,
	      as it was provided by our university anyway.
	      This statement shows that hardware does play a role and should be taken into account.
	\item \enquote{C and G are difficult to differentiate on the small laptop screen} (translated from German).
	      This goes into a similar direction as the previous statement.
	      While we checked how all tasks look on the laptop screens that are used during the exam,
	      we did not see this problem.
	      Possibly using a different font can help here.
\end{itemize}
Overall we think the results from the poll are encouraging, but there are some details that need improvement.
Especially the free text student feedback shows how we can improve the exam,
and that it is important to incorporate student feedback when designing and improving an \ac{e-exam}.

\subsection{Examiner View}\label{sec:examiner-view}

In this section we report our own experiences with the complete \ac{e-exam} workflow from creation to grading,
and highlight successes as well as difficulties.

When creating a new \ac{e-exam}, we begin with writing the tasks and solutions outside \ac{YAPS}.
Where possible, we also write tests for the solutions to ensure they are indeed correct.
During that phase we employ literate Haskell with Markdown through
\texttt{markdown-unlit}\footnote{\url{https://github.com/sol/markdown-unlit}} to combine text, code, and tests in the
same file.
This workflow has two main advantages: we can render the exam in PDF format and split between a PDF that contains all
tasks, and one that contains the solutions, and we can directly execute the exam code and tests.
For a paper exam we would instead write the exam in \LaTeX (with tests in a separate file),
which can be more time-intensive because we need to pay more attention to the resulting layout.

When the exam tasks themselves are finalized, we implement them in \ac{YAPS}.
The task texts can mostly be simply pasted into \ac{HTML} templates,
answers for single or multiple choice tasks are
easily configured in \ac{JSON} files,
and \acp{RegEx} are generated automatically by our tool.
This part of implementing the exam is a rather tedious process, but the templates and tool support help to minimize
redundancies.
Yet, we manually perform the same steps multiple times, which likely could be automated.
More creative work is required for proof and programming tasks.
Here, we need to find sensible distractors (proof tasks) and write suitable property tests (programming tasks).
This is the most time-consuming part of the entire exam creation process.
Finally, we perform a few ($< 5$) rounds of testing the finished exam to make absolutely sure everything works as
intended.
For testing, we mostly enter correct answers and check whether they are detected as such.
Usually there are few mistakes, e.g., typos, but sometimes also problems with the tests for programming tasks occur.

\begin{wraptable}{R}{.35\textwidth}
	\centering
	\caption{Comparison of paper exam (fall 2020) and \ac{e-exam} (fall 2021) results}\label{tab:resultcomp}
	\begin{tabular}{l|rr|rr}
		Task & \multicolumn{2}{r|}{Paper Exam} & \multicolumn{2}{r}{E-Exam}             \\
		     & max                             & avg                        & max & avg \\ \hline \hline
		\rowcolor{CornflowerBlue}
		1a   & 7                               & 4                          & 6   & 4   \\
		\rowcolor{RedOrange}
		1b   & 5                               & 3                          & 6   & 3   \\ \hline
		2a   & 6                               & 4                          & 6   & 4   \\
		2b   & 6                               & 3                          & 6   & 3   \\ \hline
		3a   & 10                              & 6                          & 10  & 6   \\
		\rowcolor{CornflowerBlue}
		3b   & 3                               & 1                          & 3   & 2   \\ \hline
		4a   & 6                               & 4                          & 6   & 4   \\
		\rowcolor{RedOrange}
		4b   & 4                               & 3                          & 4   & 2   \\ \hline
		\rowcolor{CornflowerBlue}
		5a   & 3                               & 1                          & 3   & 2   \\
		5b   & 5                               & 1                          & 5   & 1   \\
		5c   & 5                               & 1                          & 5   & 1   \\ \hline
		6a   & 6                               & 3                          & 6   & 3   \\
		\rowcolor{CornflowerBlue}
		6b   & 6                               & 2                          & 6   & 4   \\ \hline
		7a   & 6                               & 4                          & 6   & 4   \\
		\rowcolor{CornflowerBlue}
		7b   & 6                               & 3                          & 6   & 4   \\
		7c   & 5                               & 1                          & 5   & 1   \\ \hline
		\rowcolor{RedOrange}
		8a   & 5                               & 2                          & 6   & 1   \\
		\rowcolor{CornflowerBlue}
		8b   & 6                               & 2                          & 5   & 2
	\end{tabular}
\end{wraptable}

On the day of the exam, the required technical infrastructure is already set up by a team from our university
\cite{sitzmann2022}.
On the other hand, conducting an \ac{e-exam} may require more organizational overhead because we only have 100 laptops
available, so for large exams we need to do multiple rounds.
This introduces a new set of difficulties because we need to make sure no information about the exam is given from,
e.g., the first round to the second.
One way to do this is to provide different exams, which increases the required effort substantially.
Here, the randomization features of \ac{YAPS} are very useful to create task variations with little effort.
When time for the exam is up, \ac{YAPS} prevents any further input from students, and collects all given answers
(or rather collects them continuously during the exam to make sure no data is lost).
We can then trigger the automated evaluation, which may take some minutes, so we come back later.

Grading is then done in \ac{YAPS} as well.
We almost exclusively look at those tasks where mistakes were found by the automated grading to manually distribute points.
Additionally, we sample a few answers detected as correct to check whether the automated grading worked as intended,
which usually is the case.
If there is a mistake, however, we can correct it and execute the automated evaluation again.
Grading an exam with 77 participants took a single person approximately two work days.
The last paper exam with 136 participants took two people approximately three work days.
Unfortunately, \enquote{grading effort} is hard to measure, and the numbers are difficult to compare because of the
very different numbers of participants.
Still, it is a good sign that the exam could be graded by a single person in a comparatively short amount of time.
Assuming that with twice as many participants (154) the required amount of time would double as well,
a single person could grade as many \acp{e-exam} as two people who grade the same exam in paper, in the same timeframe.

\Cref{tab:resultcomp} shows the results per task for our last paper exam, and the \ac{e-exam} we report on in this
paper.
For each task, the maximum achievable number of points, and the average amount of points scored by students are shown.
Rows with better/worse results (in comparison to the paper exam)
are colored {\color{CornflowerBlue}blue}/{\color{RedOrange}red}.
Overall, the students performed very similarly.
The few, minor differences we see cannot be clearly attributed to a specific phenomenon.
For instance, tasks 2b, 3b, 4b, 5a, 5c, 6b, and 8a are programming tasks, but the students sometimes performed better,
worse, or the same.
When deciding for the \textit{Proof Puzzle}, we thought that the proof task (8b) may become noticeably easier
because in the paper exam the proof had to be created entirely from scratch.
However, the results suggest that in this exam creating the proof was similar in difficulty

In summary, creating an \ac{e-exam} takes more time than creating a paper exam, but with maturing templates and tools
we can confidently expect that this overhead becomes smaller.
Conducting and grading, however, takes less time now.
The exam results were almost exactly the same compared to a previous paper exam.
Altogether, we conclude that the approach to create an \ac{e-exam} from a paper exam by direct \enquote{translation} of
each task was successful.

%% file: sections/summary.tex
\section{Summary \& Future Work}\label{sec:summary}

We showed how a traditional paper exam with complex proof and programming tasks can be transformed to an \ac{e-exam}
with automated grading, but without sacrificing \ac{CA}.
Through careful analysis of the course and previous exams we can ensure that exam quality does not degrade during the
transformation.
For realizing the \ac{e-exam} we built upon existing software that we extended with a Haskell compiler, tasks that
can be checked with \acp{RegEx}, a new algorithm to automatically evaluate proofs, and a general purpose comment 
field for students.
Additionally, we introduced two new tools that substantially support automation: one that analyzes student code for
task-relevant features, and one that generates suitable \acp{RegEx} from a more high-level description language.
We achieved that almost all task can be graded automatically at least in part, and students as well as examiners
are largely satisfied with the resulting \ac{e-exam}.

In the future we want to investigate how a more fine-grained automated grading can be achieved because currently
awarding reduced amounts of points for partial solutions is often not possible
(at least not as reliably as we require it for an exam).
Moreover, because creating an \ac{e-exam} requires more effort than creating a paper exam,
we also want to explore how this process can be automated further.
Our vision is that most of the \ac{e-exam} can be generated automatically from the initial literate Haskell Markdown
file that contains the task texts, solutions, and point distribution.

%% file: sections/appendix.tex
\newpage
\appendix
\section*{Appendix}\label{sec:appendix}

\begin{figure}[H]
	\centering
	\begin{subfigure}{1.0\textwidth}
		\includegraphics[keepaspectratio, width=\textwidth]{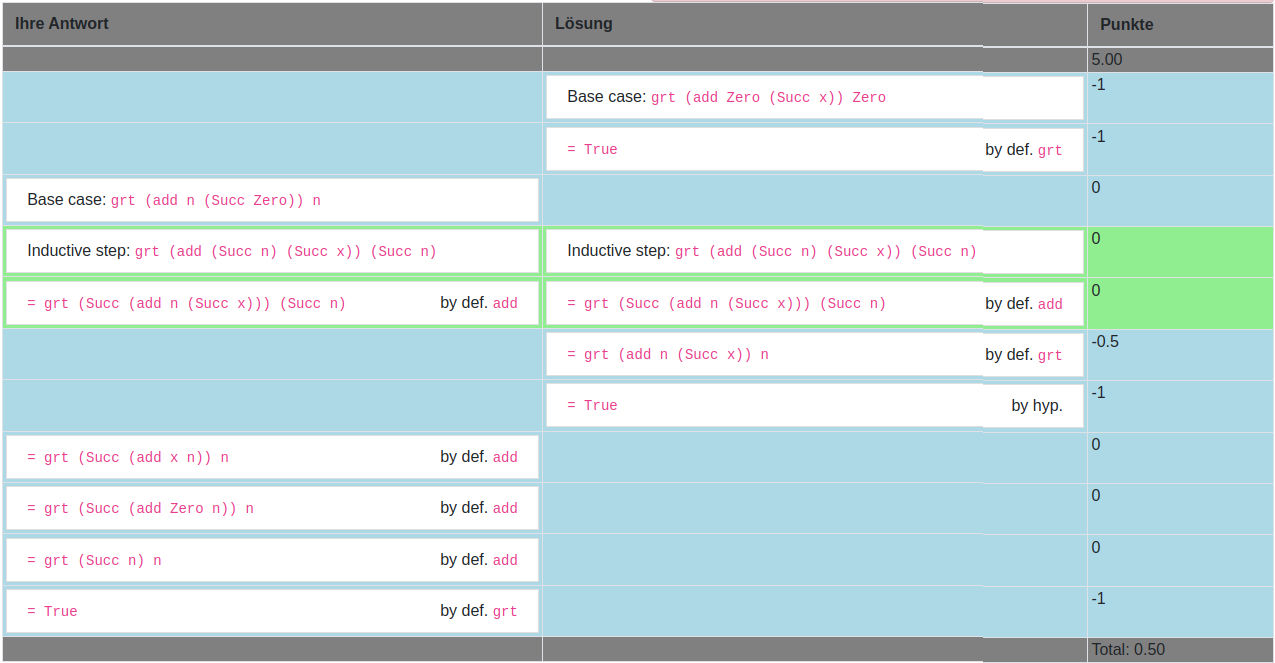}
		\caption{Student A}
	\end{subfigure}
	\begin{subfigure}{1.0\textwidth}
		\includegraphics[keepaspectratio, width=\textwidth]{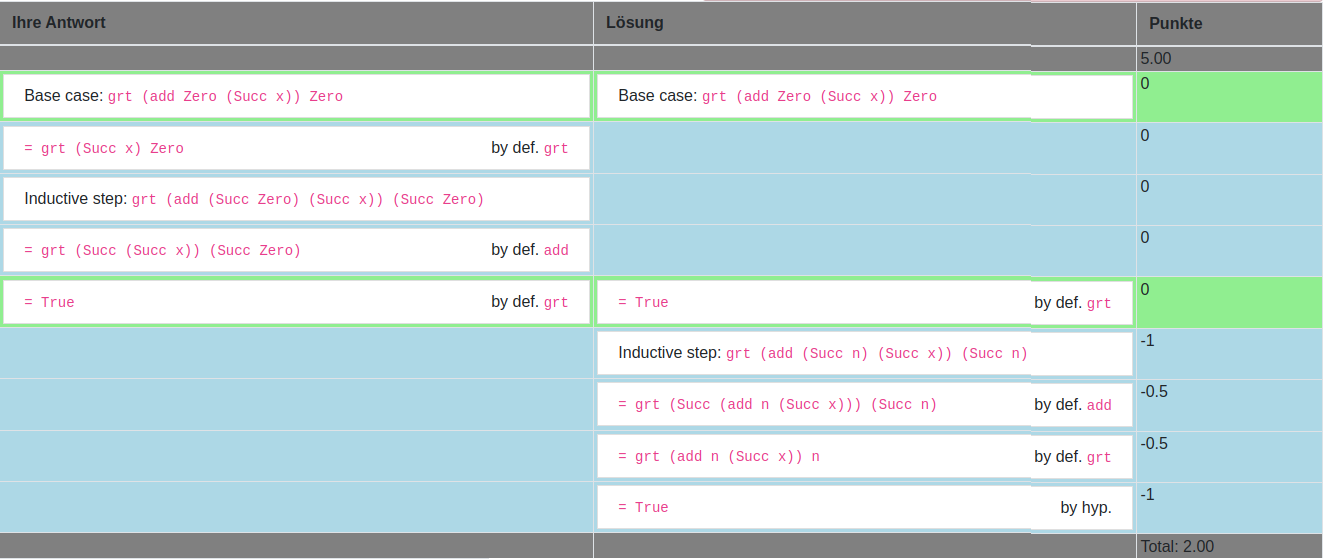}
		\caption{Student B}
	\end{subfigure}
	\caption{Example of unfair grading with old algorithm}\label{fig:proof-old}
	Left column: student solution, middle column: our solution, right column: points
\end{figure}

\begin{figure}[H]
	\centering
	\begin{subfigure}{1.0\textwidth}
		\includegraphics[keepaspectratio, width=\textwidth]{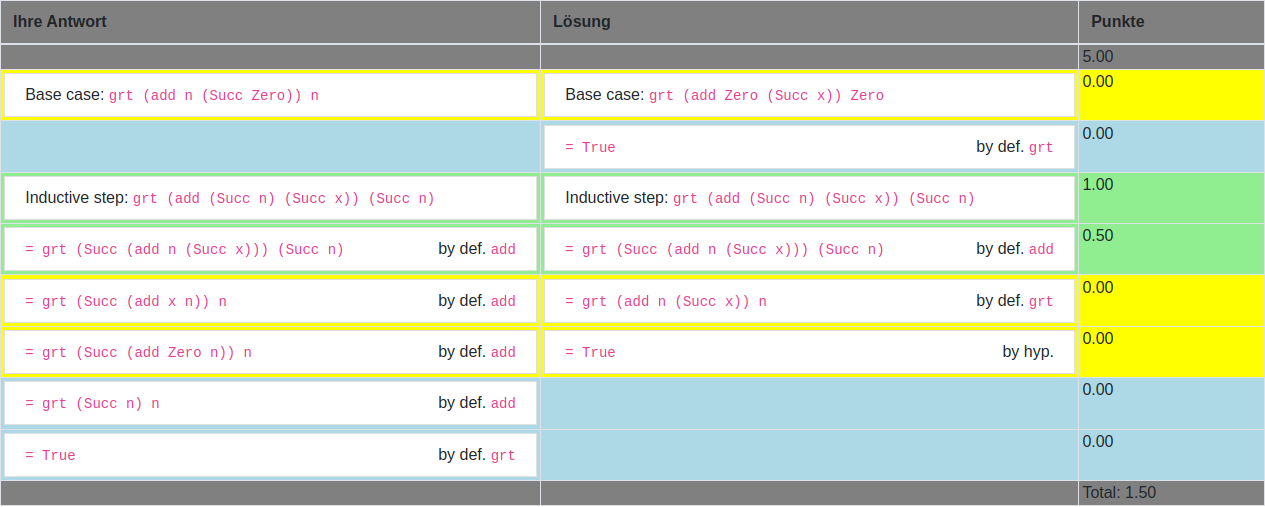}
		\caption{Student A}
	\end{subfigure}
	\begin{subfigure}{1.0\textwidth}
		\includegraphics[keepaspectratio, width=\textwidth]{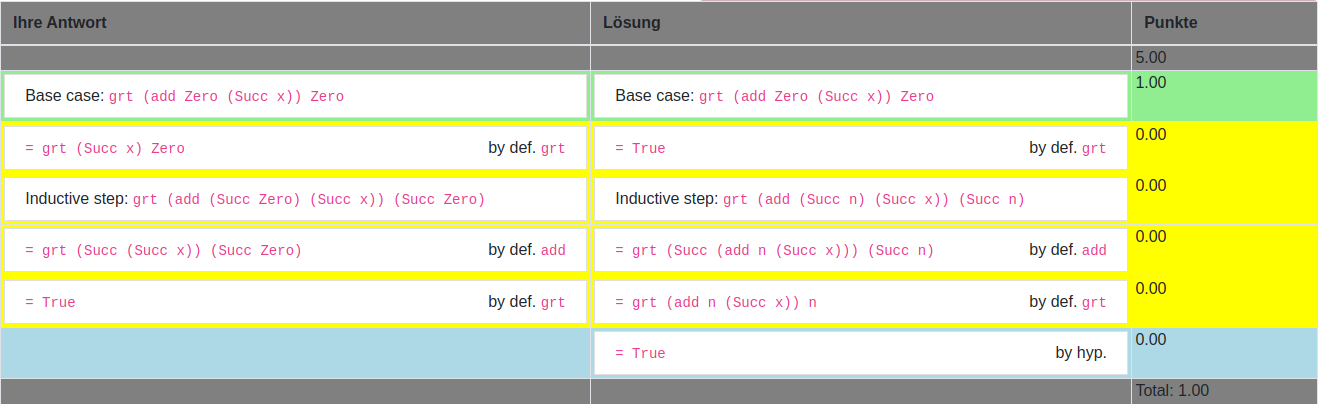}
		\caption{Student B}
	\end{subfigure}
	\caption{No unfair grading with new algorithm}\label{fig:proof-new}
	Left column: student solution, middle column: our solution, right column: points
\end{figure}

\begin{algorithm}[H]
	\caption{Default proof puzzle evaluation}\label{alg:oldproof}
	\begin{algorithmic}[1]
		\Require $I \subset \Sigma^* \times \mathbb{Q}$ (items),
		$S \subset \mathbb{S}_I \times \mathbb{Q}$ (solutions),
		$a \in \mathbb{S}_I$ (solution attempt)
		\Ensure Amount of points to award for $a$
		\State $r \gets 0$
		\ForAll{$(s_i, p_i) \in S$}
		\State $d \gets \Call{EditDistance}{s_i, a}$
		\State $r_i \gets \max\{0, p_i - d\}$
		\If{$r_i > r$}
		\State $r \gets r_i$
		\EndIf
		\EndFor
		\State \Return $r$
	\end{algorithmic}
\end{algorithm}

\begin{lstlisting}[
    morekeywords={entrypoints,separator,nonempty,comment},
    float,
    caption={\Acl{HTRSL} grammar},
    label=lst:htrsl-grammar
  ]
  entrypoints [Spec] ;
  separator Spec ";" ;

  SNoTests.   Spec ::= Desc ;

  D. Desc ::= [DescItem] ;
  separator nonempty DescItem "" ;

  DLit.   NoParensDescItem ::= String ;
  DName.  NoParensDescItem ::= Ident ;
  DSpace. NoParensDescItem ::= "\\" ;
  DNAlt.  NoParensDescItem ::= "[" [NoParensDescAlt] "]" ;
  DND.    DescItem ::= NoParensDescItem ;
  DPar.   DescItem ::= "(" [NoParensDescItem] ")" ;

  separator nonempty NoParensDescItem "" ;

  DNAltElem. NoParensDescAlt ::= [NoParensDescItem] ;
  separator nonempty NoParensDescAlt "|" ;

  comment "--";
  comment "{-" "-}";
\end{lstlisting}